\documentclass[prl,a4paper,twocolumn,11pt,accepted=2025-12-02]{quantumarticle}
\pdfoutput=1
\usepackage[utf8]{inputenc}
\usepackage[english]{babel}
\usepackage[T1]{fontenc}
\usepackage{amsmath}
\usepackage{tikz}
\usepackage{lipsum}

\usepackage{xkeyval}
\usepackage{etoolbox}
\usepackage{geometry}
\usepackage{fancyhdr}
\usepackage{tikz}
\usepackage{ltxgrid}
\usepackage{ltxcmds}
\usepackage{hyperref}
\usepackage{graphicx}
\usepackage{dcolumn}
\usepackage{bm}
\usepackage{wrapfig}
\usepackage{booktabs}

\usepackage{amsfonts}
\usepackage{amssymb}
\usepackage{graphicx}
\usepackage{bbold}
\usepackage{dsfont}
\usepackage{adjustbox}

\usepackage{accents}
\usepackage{graphicx}
\usepackage{gensymb} 
\usepackage{dcolumn}
\usepackage{bm}
\usepackage{amsmath}
\usepackage{graphicx}
\usepackage{xcolor}
\usepackage{float}
\usepackage{mathtools}
\usepackage{comment}
\usepackage[normalem]{ulem} 
\usepackage{subfigure}
\usepackage{changepage}

\DeclareMathOperator{\tr}{Tr}
\newcommand{\id}{\mathds{1}}
\usepackage{amsthm}

\newcommand{\ket}[1]{\vert {#1} \rangle}

\newcommand{\Bra}[1]{{ \langle \! \langle{#1}\vert }}
\newcommand{\Ket}[1]{{ \vert {#1}  \rangle \!  \rangle}}
\newcommand{\KetBra}[1]{{\Ket{#1}\!\Bra{#1} }}

\newcommand{\blockcomment}[1]{} 

\begin{document}

\title{Multi-time quantum process tomography on a
superconducting qubit}

\author{Christina Giarmatzi}
\thanks{These authors contributed equally to this work\\
christina.giar@gmail.com\\
tyler.jones@fiasqo.io}
\affiliation{School of Computer Science, University of Technology Sydney, Ultimo, Sydney, New South Wales 2007, Australia}
\affiliation{ARC Centre of Excellence for Engineered Quantum Systems, St. Lucia, Brisbane, Queensland 4072, Australia}
\affiliation{
 School of Mathematical and Physical Sciences, Macquarie University, Sydney, New South Wales 2122, Australia
}

\author{Tyler Jones}%
\thanks{These authors contributed equally to this work\\
christina.giar@gmail.com\\
tyler.jones@fiasqo.io}
\affiliation{ARC Centre of Excellence for Engineered Quantum Systems, St. Lucia, Brisbane, Queensland 4072, Australia}
\affiliation{%
 School of Maths and Physics, University of Queensland, St. Lucia, Brisbane, Queensland 4072, Australia
}
\affiliation{%
 Fiasqo, Brisbane, Queensland 4072, Australia
}%

\author{Alexei Gilchrist}
\affiliation{ARC Centre of Excellence for Engineered Quantum Systems, St. Lucia, Brisbane, Queensland 4072, Australia}
\affiliation{
 School of Mathematical and Physical Sciences, Macquarie University, Sydney, New South Wales 2122, Australia
}%
\author{Prasanna Pakkiam}
\affiliation{ARC Centre of Excellence for Engineered Quantum Systems, St. Lucia, Brisbane, Queensland 4072, Australia}
\affiliation{%
 School of Maths and Physics, University of Queensland, St. Lucia, Brisbane, Queensland 4072, Australia
}
\author{Arkady Fedorov}
\affiliation{ARC Centre of Excellence for Engineered Quantum Systems, St. Lucia, Brisbane, Queensland 4072, Australia}
\affiliation{%
 School of Maths and Physics, University of Queensland, St. Lucia, Brisbane, Queensland 4072, Australia
}
\author{Fabio Costa}
\affiliation{%
 School of Maths and Physics, University of Queensland, St. Lucia, Brisbane, Queensland 4072, Australia
}

\affiliation{Nordita, Stockholm University and KTH Royal Institute of Technology, Stockholm, 106 91, Sweden}

\begin{abstract}
Current quantum technologies are at the cusp of becoming useful, but still face formidable obstacles such as noise.
Noise severely limits the ability to scale quantum devices to the point that they would offer an advantage over classical devices. 
To understand the sources of noise it is necessary to fully characterise the quantum processes occurring across many time steps; only this would reveal any time-correlated noise called non-Markovian.
Previous efforts have attempted such a characterisation but obtained only a limited reconstruction of such multi-time processes. 
In this work, we fully characterise a multi-time quantum process on superconducting hardware using in-house and cloud-based quantum processors. 
We achieve this by employing sequential measure-and-prepare operations combined with post-processing. 
Employing a recently developed formalism for multi-time processes, we detect general multi-time correlated noise. 
We also detect quantum correlated noise which demonstrates that part of the noise originates from quantum sources, such as physically nearby qubits on the chip. 
%
%
\end{abstract}

\maketitle


\section{\label{sec:level1}Introduction}

The characterisation of noise is a critical requirement for the advancement of quantum technologies. Noise is present in all current quantum devices as they interact with their environment; for many devices, this includes non-Markovian noise due to temporal correlations across a multi-time process. While most current techniques assume Markovian noise, this assumption fails in realistic quantum devices and non-Markovian noise has been reported in state-of-the-art quantum computing devices (such as IBM and Google)~\cite{Morris2022, McEwen2021,Sandia_report}, {having a detrimental effect on computation and fault-tolerance thresholds~\cite{Nickerson2019analysingcorrelated}}.

{Non-Markovian noise is essentially correlated noise across time. In quantum computing, this can appear as correlated errors between gates of a quantum circuit occurring at different times on the same system. This leads to a noisy quantum process, even if all its elements have been individually optimised against noise. A non-Markovian process can be simulated by adding an environment that interacts with the system at different times and goes through a channel in-between. Non-Markovian noise can be divided into classical and quantum, depending on the nature of the channel needed to simulate it; quantum or classical channel. Classical non-Markovian noise can originate, for example, from slowly drifting environmental fluctuations such as noise from electronics or nearby classical or quantum systems. Quantum non-Markovian noise requires a quantum system to be the environment that correlates the different system-environment interactions in time.}

{Traditional} approaches to capture non-Markovian noise only provide access to two times of a quantum process through dynamical maps~\cite{Piilo08, Breuer09, Rivas10, Rivas2014, Breuer16,Chruscinski2011, Vega:2017aa}. Beyond this, a number of {experimental} techniques designed to reduce non-Markovian noise have been developed, {especially within spin silicon qubits}~\cite{takeda18,Freer:2017aa,Savytskyy:aa,Zwerver:2022aa,Philips:2022aa}, and shown to increase computational accuracy~\cite{takeda18}, but all rely on ad-hoc methods {that are not based on principled characterisation of non-Markovian processes.}

Recently, a formalism was developed that constructs a \emph{process matrix}~\cite{oreshkov12,oreshkov15}, which encodes all multi-time correlations in a quantum process. In this formalism, non-Markovian noise is rigorously captured and can be further investigated in terms of its nature (quantum or classical) and amount~\cite{Giarmatzi2021, Shrapnel2018}, as well as memory length and location in the device~\cite{taranto2021} with a prospect of scalability~\cite{PhysRevLett.126.200401} and extending current characterisation methods to non-Markovian regimes~\cite{RBMnmnoise}. The approach has been used in distinct experimental scenarios, successfully detecting non-Markovian noise~\cite{Morris2022,Milz2018,taranto2021,munoz2022,Goswami2021,White2020}. 

However, full reconstruction of a multi-time process matrix requires implementing sequential informationally-complete operations, such as measure-and-prepare, on the system at \emph{every} time step. Based on the assumption that this procedure requires mid-circuit measurements with a fast feed-forward of signal---unavailable in current quantum computing platforms---previous work did not succeed in performing full tomography but rather achieved a partial characterisation of non-Markovian noise, for example through a `restricted' process matrix{, an object containing partial information compared to the full process matrix}~\cite{White2020, Liang2021, White2025whatcanunitary, White2022, Li_2024, PhysRevLett.134.010803}. 

Here, {we implement the first full tomography of a multi-time quantum process on a superconducting qubit and obtain a complete description of its non-Markovian noise. We achieve this through mid-circuit measurements and a {novel} post-processing {technique} that does not require a feed-forward mechanism. This allows us to obtain complete data for full process tomography; something that has not been achieved before.} We use devices from the labs of the University of Queensland and the cloud capabilities of IBM Quantum~\cite{IBMQ}.
We measure general and quantum non-Markovian noise in all cases and build a theoretical model to compare our findings. {Our code and data are publicly available~\cite{tomo_code}.} Our work offers a robust method for {complete characterisation of} non-Markovian noise, {through the reconstruction of the full process matrix.}

\section{\label{sec:level2}Multi-time quantum process tomography}

\begin{figure}[ht]
    \centering
    \includegraphics[scale=.16]{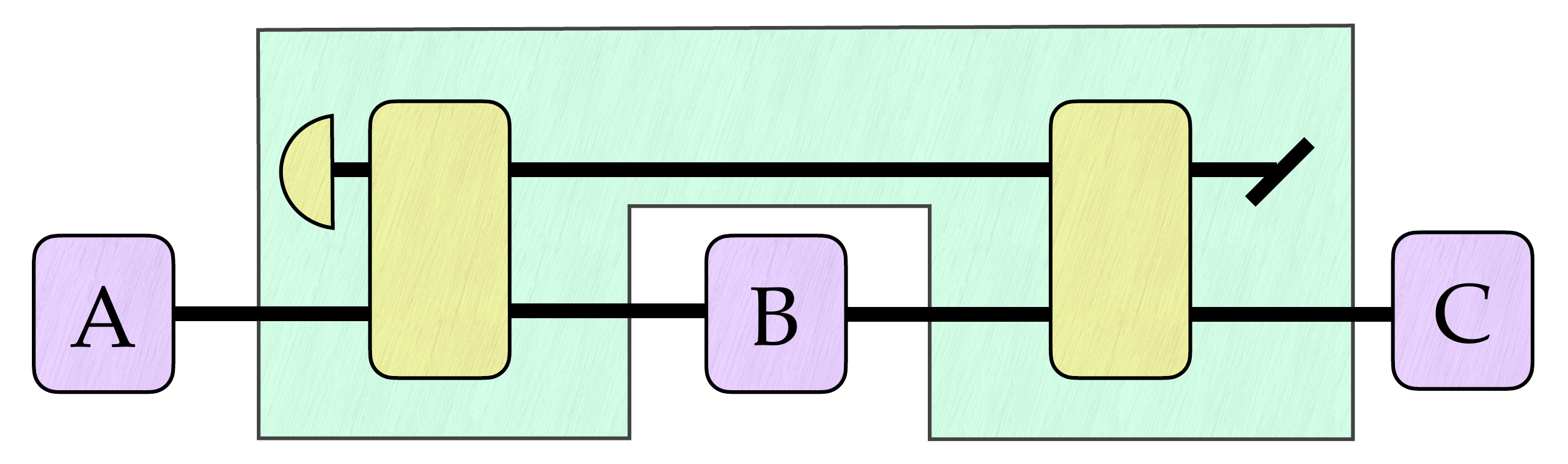}
    \caption{A three-time quantum process involving state preparation and two gates. The bottom line is the system and the top line is the environment. The boxes $A, B$ and $C$ denote operations on the system at different times.}
    \label{fig:process}
\end{figure}

Consistent with our experimental setup, we consider a scenario where a single quantum system undergoes a sequence of three operations at specified times, see Fig~\ref{fig:process}. The purpose of multi-time process tomography is to extract, from a finite set of measurements, sufficient information that allows one to predict the joint probability for any other sequence of measurements \cite{chiribella09b, costa2016, White2022}. 
The mathematical object that encodes all needed information is the process matrix---also known as quantum strategy~\cite{gutoski06}, comb~\cite{chiribella09b}, or process tensor~\cite{Pollock:2018aa}---which is a convenient representation of a quantum stochastic process~\cite{Lindblad1979}. 

Let us label $A, B,$ and $C$ the times at which the operations take place. In Figure~\ref{fig:process}, the bottom line is the system and the top line is the environment. The operations are represented by general completely positive (CP), trace-non-increasing maps from an input to an output Hilbert space so that, for example at time $A$, the joint input-output space is $\mathcal{H}^{A}=\mathcal{H}^{A_I} \otimes \mathcal{H}^{A_O}$. The map can be expressed in Choi form~\cite{jamio72,Choi1975} (see Supplementary Material) as an operator $M^{A}_{a\mid x}\in \mathcal{L}(\mathcal{H}^{A})$, $M^{A}_{a\mid x}\geq 0$, where $a$ denotes the measurement outcome, $x$ denotes the setting (i.e., the choice of operation), and $\mathcal{L}(\mathcal{H})$ the space of linear operators over $\mathcal{H}$ \cite{Heinosaari2011}. For each setting choice $x$, a complete set of outcomes defines an instrument~\cite{Davies1970}, expressed as a set $\{M^{A}_{a\mid x}\}_a$ such that $\sum_a \tr_{A_O}M^{A}_{a\mid x} = \id^{A_I}$. The process matrix is an operator on the joint space of all inputs and outputs, $W^{ABC}\in \mathcal{L}(\mathcal{H}^{A}\otimes \mathcal{H}^{B} \otimes \mathcal{H}^{C})$, $W^{ABC}\geq 0$, which allows one to calculate arbitrary joint probabilities for a sequence of outcomes $a,b,c$, given settings  $x,y,z$, through the generalised Born rule~\cite{Shrapnel_2018}
\begin{equation}
    p(a,b,c\mid x,y,z) =  
    \tr[W^{ABC} (M_{a\mid x}^{A} \otimes M_{b\mid y}^{B}\otimes M_{c\mid z}^{C})].
    \label{eq:Born}
\end{equation}

Formally, we can regard $W^{ABC}$ as a ``density operator over time'', where each time instant corresponds to a pair of subsystems, an input and an output space. As output spaces encode the causal influence from an operation to future ones, the final output space can always be discarded (hence, the final time $C$ has only an input space). Furthermore in our scenario the first operation is a strong state preparation, which decorrelates the system from the environment allowing us to discard the first input space $A_I$. Thus for our experiment the process matrix is $W^{ABC}\equiv W^{A_OB_IB_OC_I}$.

One can reconstruct the process matrix using similar techniques as for ordinary state tomography, namely one can invert Eq.~\eqref{eq:Born} and write $W$ as a function of the probabilities. A convenient choice for the instruments is projective measurements $\{\Pi_{a\mid x}\}_a$ immediately followed by the preparation of a state $\rho_x$, which correspond to operations of the form $M_{a\mid x}^{A}=\Pi_{a\mid x}^{A_I}\otimes\rho_x^{A^T_O}$, where $A^T_O$ denotes transposition on system $A_O$ and reflects our choice of convention, consistent with Eq.~\eqref{eq:Born}. {At first, it might seem that this type of measure-re-prepare instrument requires a very fast information feed-forward and choice of setting: After an ordinary projective measurement, the system is left in a state that depends on the measurement outcome. In order to re-prepare a previously chosen state, one needs to implement a unitary that depends on the outcome. In typical devices,} {measure-re-prepare instruments relying on conditional gates require specialised hardware and are not achievable on all systems, due to data processing constraints.} {This specific point has been a bottle neck in all previous work on experimental multi-time process tomography, which resorted to partial reconstruction techniques, based either on sequences of unitaries with a final measurement~\cite{PhysRevLett.134.010803, Li_2024, White2025whatcanunitary} or projective measurements without independent repreparation \cite{Liang2021}. In the next section, we present a method that overcomes this limitation without any additional technological requirement.}

\section{Multi-time tomography protocol on a superconducting qubit}
The experiment was conducted separately on two sets of superconducting hardware; an in-house processor at the University of Queensland, which will be referred to as \texttt{uq}, and a cloud processor from IBM Quantum with the moniker \texttt{ibm\_perth}. 

The experiment conducted on the \texttt{uq} processor used two capacitively-coupled flux-tunable transmons on a 5-qubit chip, mounted in a dilution refrigerator with a base temperature of around 15 mK. Each qubit is coupled to a dedicated resonator for readout. Flux lines are used in conjunction with a magnetic coil fixed to the chip to control transition frequencies, providing a degree of freedom by which qubit-qubit interaction strengths can be controlled.

\begin{figure*}[!htbp]
    \centering
    \includegraphics[width=0.85\textwidth]{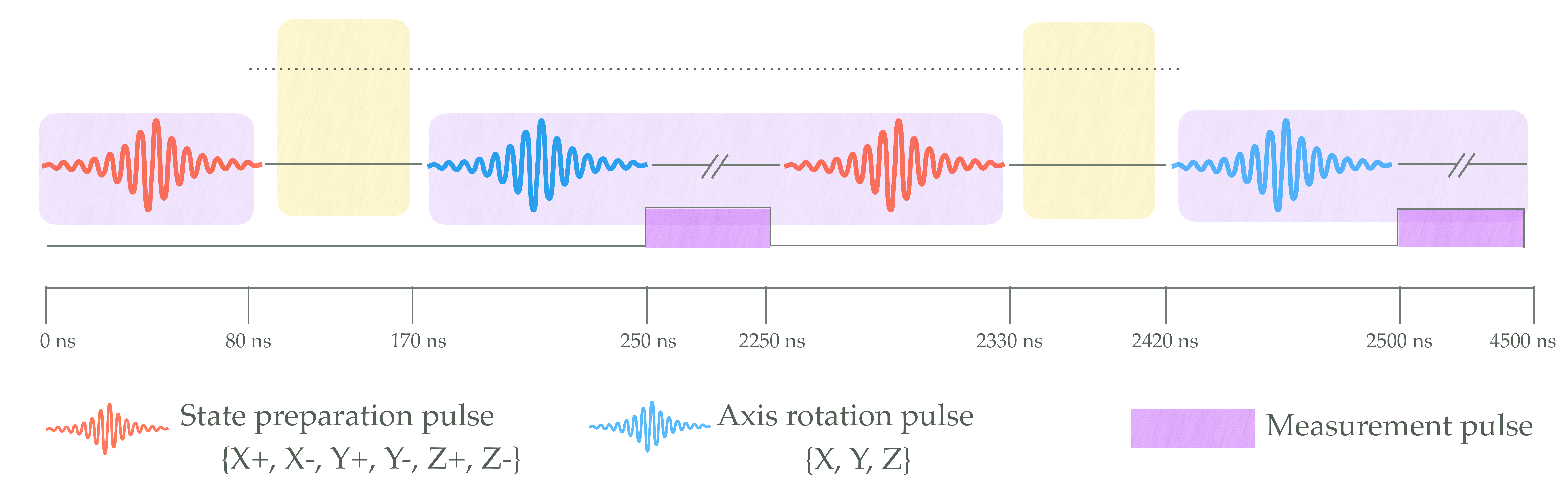}
    \caption{A visualisation of the pulse protocol for the tomography data measured on the \texttt{uq} system. The solid and dotted line is the system and environment respectively, and the two yellow boxes are the system-environment interaction.}
    \label{fig:protocol}
\end{figure*}

The experiments on the \texttt{ibm\_perth} processor addressed a single qubit on an 8-qubit chip. In both experiments, the system of interest is a single qubit chosen to be a corner qubit on the chip such that its closest environment is a single nearby qubit, a case that is studied with model simulations in the Supplemental Material. In general, though, all the other qubits are part of the environment which also comprises any other degrees of freedom interacting with the system, as well as classical sources of noise.

To implement the three-time process tomography, we set the Pauli operators as our observables $\mathcal{M} = \{X, Y, Z\}$ and their eigenstates as the input states $\mathcal{B} = \{X_+, X_-, Y_+, Y_-, Z_+, Z_-\}$. These form a tomographically (over) complete basis. We also define a set of operations within a basis gate set $\mathcal{G} = \{I, R_X(\frac{\pi}{2}), R_X(-\frac{\pi}{2}), R_Y(\frac{\pi}{2}), R_Y(-\frac{\pi}{2}), R_X(\pi)\}$. These operations are, by necessity, sufficient to prepare and measure all combinations of basis states and observables. Multi-time process tomography consists of temporally stacking two traditional process tomography protocols, forming a (prepare $\rightarrow$ process $\rightarrow$ measure $\rightarrow$ prepare $\rightarrow$ process $\rightarrow$ measure) protocol, where preparations are on spaces $A_O, B_O$, while measurements are on $B_I, C_I$. In these experiments, no operation is performed between each preparation and the subsequent measurement, so the process is composed of two identity channels from $A_O$ to $B_I$ and from $B_O$ to $C_I$. Hence, by `process' in the above protocol, we mean that for times $t_1$ and $t_2$ the system freely evolves and so the ideal channels realised should be the identity channels. Tomography aims to characterise departures from this idealisation, representing noise in the process due to a system-environment interaction during the free evolution. The preparation states are drawn from $\mathcal{B}$ and measurement observables from $\mathcal{M}$, resulting in $\lvert\mathcal{B}\rvert^2\lvert\mathcal{M}\rvert^2 = 324$ unique protocols. The general form of this protocol is illustrated in Fig. \ref{fig:protocol}. Together with Fig. \ref{fig:process}, we can see the multi-time quantum process and the times when we make the operations to gather the data. Note that the system can in principle interact with its environment at any time, i.e., during the tomography operations too. {The tomography protocol does not capture this interaction nor the potential non-Markovian dynamics arising at the timescales shorter than the measurement pulse at $B$. However, as we show in this work, there is non-Markovian dynamics to be observed at larger timescales. In addition, these limitations are expected to diminish with the development of hardware to allow for shorter measurement pulses, or with alternatives to the mid-circuit measurements.} For example, new developments claim to have achieved characterisation of the noisy tomography operations as well as the process~\cite{white2023unifying}.

To begin, all qubits were initialised in the ground state through a standard long wait-time routine. This is a deterministic operation and ensures that both system and environment are decorrelated. The system qubit is then prepared into some basis state $b_0 \in \mathcal{B}$ using the appropriate rotation $p_0 \in \mathcal{G}$. This occurs at $A$. The system and environment are then left to freely evolve for time $t_1$.

At $B$ we apply a projective measurement of the observable $m_0 \in \mathcal{M}$ on the system qubit, consisting of some rotation gate $r_0 \in \mathcal{G}$ to facilitate measurement of the correct axis, followed by a readout signal. The reflected readout signal is then classified as 0 or 1 and the result is stored in memory. After the mid-circuit measurement, another rotation $p_1 \in \mathcal{G}$ is applied to prepare a new qubit state. Conditional on the eigenstate collapse in the mid-circuit measurement, this operation prepares the system qubit into one of the basis states $b_1, \tilde{b}_1 \in \mathcal{B}$. For example, if a protocol measures $m_0 = X$ and applies rotation $p_1 = I$, the prepared state is either $b_1 = X_+$ or $\tilde{b}_1 = X_-$. The actual prepared state is then identified by checking the result of the mid-circuit measurement. In this way, we can obtain a complete set of preparations over all experimental runs without the need for each prepared state to be chosen deterministically. This simple but critical post-processing technique is designed to avoid the need for dynamic, fast feed-forward control, which would otherwise be required for conditional operations. Following these operations, the system and environment are left to freely evolve for time $t_2$. In this protocol, we chose $t_1$ and $t_2$ according to our simulations (see Supplemental Material). Assuming one more qubit as the environment, the given experimental parameters of each qubits resonance frequency, their interaction parameter and their Hamiltonian, we obtained several pairs of $(t_1,t_2)$ that maximise non-Markovianity of the process.

Finally, at $C$ we measure our second observable $m_1 \in \mathcal{M}$. In all, the protocol is repeated $2^{14}$ times for each of 324 prepare-measure-prepare-measure permutations, resulting in approximately $5.3\times10^{6}$ individual shots per process characterisation experiment (one on \texttt{uq} and nine at \texttt{ibm\_perth}). This data is shaped into the number of realised counts we obtain for each permutation. An example is $(X_+, X_-, Y_+, Z_-)$ denoting that at $A$ we prepare the state $X_+$, at $B$ we projectively measure $X_-$ and then re-prepare $Y_+$, and at $C$ we projectively measure $Z_-$.

After obtaining the data, we are ready to calculate the experimental process matrix. A process matrix can written in the Pauli basis, $W = \sum_{i,j,k,l}\alpha_{i,j,k,l}\sigma_i\otimes \sigma_j\otimes \sigma_k\otimes\sigma_l$, where (i,j,k,l) = [0,3] denotes the Pauli basis ${\id,\sigma_x,\sigma_y,\sigma_z}$. From Eq. \ref{eq:Born}, we see that if we apply Pauli operations at $A,B,C$, their expectation values will be the coefficients $\langle\sigma_i,\sigma_j,\sigma_k,\sigma_l\rangle = \alpha_{i,j,k,l}$. Hence, we can reconstruct the experimental process matrix, by calculating the expectation values of the sigma terms from the experimental data
\begin{equation} \label{inversion}
    W_{exp} = \frac{1}{16}\sum_{i,j,k,l}\langle\sigma_i,\sigma_j,\sigma_k,\sigma_l\rangle\sigma_i\otimes\sigma_j\otimes\sigma_k\otimes\sigma_l.
\end{equation}
$W_{exp}$ is the process matrix we construct from the experimental data, but it typically does not represent a physical one (it might be non-positive, or fail the process matrix constraints). {This is typically also the case for quantum state tomography, and various methods are used to obtain the physical state from the experimental one, such as maximum likelihood estimation.} To obtain a physical process, $W_{phys}$, we use SemiDefinite Programming (SDP), where we find a positive matrix that minimises the distance (we used Frobenius distance) to $W_{exp}$. The SDP is the following:
\begin{align}
    \texttt{variable}&\ \ \texttt{$W_{phys}$ Hermitian Semidefinite} \nonumber \\
    \texttt{minimize}&\ \ \texttt{Frobenius Norm}(W_{exp}-W_{phys})\\
    \texttt{subject to}&\ \  L_V(W_{phys}) == W_{phys} \nonumber\\
    &\ \ \text{Tr}(W_{phys}) == 4, \nonumber
\end{align}
where $L_V$ is an operator that projects a matrix to the space of valid process matrices~\cite{araujo15},
\begin{align} \label{Lv}
    L_V(W) = W + _{C_I}W - _{B_OC_I}W + _{B_IB_OC_I}W\nonumber\\
    - _{A_OB_IB_OC_I}W, \\
    _XW = \tr_{X}(W\otimes\id^X)/d_X, \nonumber
\end{align}
and the normalisation constraint helps the SDP to bound the solution. Using the $W_{phys}$ we report in the Results section our findings on tomography and non-Markovianity. {Note that the step from a non-physical process matrix to a physical one is a well-known problematic step in state tomography, because it does not allow propagation of SPAM {(state preparation and measurement)} errors from the data to the physical process or state. Prior works have extensively used Maximum Likelihood Estimation (MLE) for this step, specifically to go from non-informationally-complete data to a physical process. Both SDP and MLE methods have been shown to perform well in process tomography tasks and has been shown that for a sample size of $\approx 10,000$ (as in our case of $8,000$ shots) CVX and ML performed comparably in~\cite{PhysRevA.98.062336}. However, both methods, and other similar attempts, cannot propagate SPAM errors. For this reason, in the Appendix we address the sampling and measurement error and obtain error bars to our reported results.} {Further improvements may be developed in future work by translating to multi-time process tomography methods developed for states, for example adapting gate set tomography to account for SPAM errors \cite{Nielsen2021gatesettomography}.}

Our tomography protocol differs from related works in that we implement full multi-time process tomography by reconstructing the full process matrix. Concretely, Xiang \emph{et al.} implement projective mid-circuit measurements but limit their analysis to an under-complete set of POVMs to construct a `restricted' process matrix~\cite{Liang2021}. Similarly, White \emph{et al.} restrict their tomography to only sequences of unitary operations and final measurement, with reference to the difficulty of mid-circuit measurement in superconducting processors~\cite{White2022}. Related work by the same authors extends their analysis to non-unital maps by introducing an ancillary qubit and entangling operations, obtaining a separate class of restricted information known as classical shadows~\cite{White2025whatcanunitary}. {Hence, prior works have obtained non-complete information about the process (restricted process or classical shadows), offering a partial characterisation of its non-Markovianity.} {In this work, we obtain} the full process matrix and provide the first complete characterisation of non-Markovian noise, which underpins the novelty of our method and findings.

\section{Results: Tomography and non-Markovian noise}
\begin{figure*}[!htbp]
    \centering
    \includegraphics[width=0.6\textwidth]{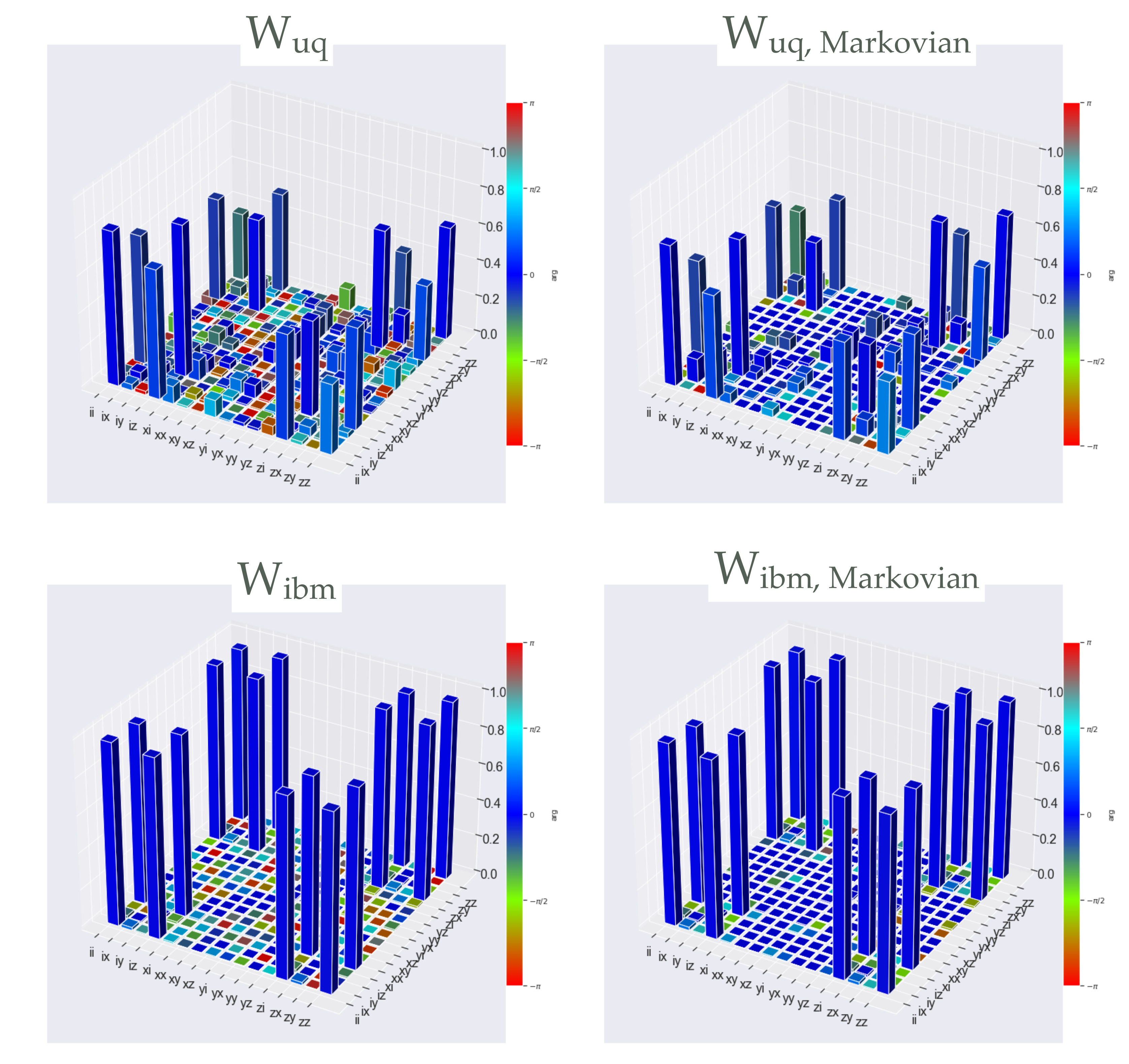}
    \caption{A visualisation of the absolute valued entries ($z$ axis) with the phase as colour, of the full 16x16 multi-time process matrix ($x$ and $y$ axis are running across the 2 dimensions of the matrix),  constructed with tomographic data from the \texttt{uq} and the \texttt{ibm\_perth} processor. The adjacent subfigures depict the closest Markovian process matrix.}
    \label{fig:matrices}
\end{figure*}

Figure \ref{fig:matrices} provides insight into the qualitative differences between two experimental (physical) process matrices, $W_{phys}$, using data from the in-house \texttt{uq} processor and from the \texttt{ibm\_perth} processor. 
The process matrix from \texttt{ibm\_perth} is visually almost indistinguishable from a noiseless (or Markovian) process with identity channels between the times $A$, $B$ and $C$, while the \texttt{uq} process maintains the qualitative structure of the identity channels with evident noise. 

Overall, we took tomographic data for one process on the \texttt{uq} processor and nine on \texttt{ibm\_perth} with varying $t_1$ and $t_2$. Variations in free evolution times were explored after simulating each system with a simple one-qubit environment model and finding that both general and quantum non-Markovianity depend highly on these times, with a particular structure (see Supplementary Material). While we chose the experimental $t_1,t_2$ that maximise non-Markovianity in the simulation, no such structure was observed in the experiments we conducted.

Given the physical process matrices for each of the ten processes characterised (broadly denoted as $W_{phys}$), we aim to detect and quantify non-Markovianity. To this end, we calculate a measure of distance of each $W_{phys}$ with its closest Markovian process matrix $W_{M}$.
We construct $W_{M}$ by discarding temporal correlations, combining the two channels ~\cite{White2022,Giarmatzi:2018aa}
\begin{align}
    W_{M} = T_1^{A_OB_I}\otimes T_2^{B_OC_I},
\end{align}
where $T_1^{A_OB_I} = Tr_{B_OC_I}W_{phys}$ and $T_2^{B_OC_I} = Tr_{A_OB_I}W_{phys}$.

{As a measure of distance, we consider the relative entropy for two matrices $W_1$ and $W_2$ (normalised as states with trace 1),}
\begin{equation}
    \mathcal{D}(W_1||W_2) = \tr(W_1\log_2W_1 - W_1\log_2W_2),
    \label{eq:s}
\end{equation}
where $W_1 = W_{phys}/\tr{W_{phys}}$ and $W_2 = W_{M}/\tr{W_{M}}$. {$\mathcal{D}$ is 0 for equal matrices and is unbounded from above.} 
{Note that the quantum relative entropy determines the asymptotic probability of distinguishing two states \cite{1991hia99,2002ved197}, and this meaning carries over to process matrices.}
The results for $\mathcal{D}$ across the ten processes are in Table \ref{table:metrics}. {A non-zero value indicates non-Markovian noise} in both platforms for all choices of parameters {but stronger on the \texttt{uq} processor}. This is consistent with the appreciably stronger idle qubit-qubit interaction on this chip.

\begin{table}[htt]
\begin{adjustbox}{width=.85\columnwidth,center}
\begin{tabular}{ccc} \toprule
      \small{Evolution} & \small{General} & \small{Quantum} \\ \midrule
     {($t_1, t_2$)} ns & {$\mathcal{D}$} & {${\mathcal{N}}$} \\ \midrule
     (97, 97)  & $0.2318 \pm 0.0080$ & $0.0215 \pm 0.0024$  \\ \midrule
     (21.333, 21.333)  & $0.0446 \pm 0.0074$ & $0.0043 \pm  0.0015$ \\
     (21.333, 24.889)  & $0.0382 \pm 0.0070$ & $0.0044 \pm 0.0014$ \\
     (21.333, 28.444)  & $0.0471 \pm 0.0075$ & $0.0057 \pm 0.0017$ \\
     (24.889, 21.333)  & $0.0364 \pm 0.0068$ & $0.0053 \pm 0.0016$ \\
     (24.889, 24.889)  & $0.0467 \pm 0.0070$ & $0.0049 \pm 0.0014$ \\
     (24.889, 28.444)  & $0.0579 \pm 0.0073$ & $0.0065 \pm 0.0015$ \\
     (28.444, 21.333)  & $0.0540 \pm 0.0074$ & $0.0055 \pm 0.0015$ \\
     (28.444, 24.889)  & $0.0462 \pm 0.0069$ & $0.0059 \pm 0.0013$ \\ 
     (28.444, 28.444)  & $0.0421 \pm 0.0070$ & $0.0048 \pm 0.0014$ \\ \bottomrule \\
\end{tabular}
\end{adjustbox}
\caption{The {relative entropy ($\mathcal{D}$}) and negativity ($\mathcal{N}$) of the multi-time process matrices reconstructed for each processor. Each experiment has two periods of free evolution, $t_1$ and $t_2$. The first row shows the \texttt{uq} data and the next nine runs were from \texttt{ibm\_perth}. {The error bars, obtained in the Appendix, account for sampling and measurement errors.}}
\label{table:metrics}
\end{table}

A natural subsequent inquiry is whether non-Markovianity originates from a quantum source; the value of distinguishing between classical or quantum environments {or memory} lies in the manifestly different dynamics they induce~\cite{Jouzdani2014,PazSilva2017,Beaudoin2018,PazSilva2019}. {This can be described as the need for a classical or quantum channel to simulate the multi-time process.} The quantum nature of non-Markovianity in the process is captured by quantum correlations across the evolution from $A$ to $B$ and from $B$ to $C$~\cite{Giarmatzi2021}. This maps to entanglement across the bipartition $A_OB_I\vert B_OC_I$ in the four-qubit state representation of the process, $\rho^{A_OB_I\vert B_OC_I} := {W}/{\tr W}$ . {A simple entanglement criterion is the well-known Positive Partial Transpose (PPT) criterion, which states that if the partial transpose of a state is negative, then the state is entangled. This criterion yields an easily computable measure of entanglement called the \emph{negativity},}
\begin{equation}
    \mathcal{N}(\rho)=\left\lvert\sum_{\lambda_i<0} \lambda_i\right\rvert=\sum_i \frac{\lvert\lambda_i\rvert-\lambda_i}{2},
    \label{eq:neg}
\end{equation}
where {$\rho=\rho^{(A_OB_I)^T\vert B_OC_I}$}, where {$T$ denotes partial transposition (it can be on either of the two subsystems, $A_OB_I$ or $B_OC_I$), and} $\lambda_i$ are the eigenvalues of $\rho$~\cite{Vidal2002}. {Any value strictly larger than zero is a signature of entanglement.} In state space, negativity is upper bounded by the maximally entangled state ($\frac{3}{2}$ for four qubits), roughly two orders of magnitude larger than in our measurements. However, the process matrix space is smaller than the state space due to the extra causality constraints, {and an operational meaning is still an open problem.}

Quantum multi-time correlations are observed on each processor, but are markedly enhanced on the $\texttt{uq}$ processor. This is consistent with the increased interaction strength observed between the system qubit and its primary interacting `memory' qubit on the $\texttt{uq}$ processor. The negativities reported over the nine $\texttt{ibm\_perth}$ runs all lay within each others' credible intervals, indicating that the quantum non-Markovianity present in each process was effectively invariant across the ($t_1$, $t_2$) landscape explored. In all measurements, the reported uncertainty denotes 95\% credible intervals determined by bootstrapping techniques (in the Appendix), accounting for sampling noise {as well as experimental measurement noise}. 


In summary, we observe and quantify non-Markovian {general and quantum} dynamics with high confidence on both $\texttt{uq}$ and $\texttt{ibm\_perth}$ processors, highlighting the relevance of correlated noise in the design of future mitigation techniques. Our results demonstrate the practicality of our methods for characterising this noise.\\

\section{Conclusion}\label{sec5}
We report the first experimental full {tomography} of a multi-time quantum process on a superconducting qubit. {We implement mid-circuit measurements and overcome the need for a feed-forward mechanism with a {new} post-processing {technique}.} We implement ten three-time processes using an in-house superconducting processor and a cloud processor from IBM Quantum. We characterise the processes using the process matrix formalism and provide a measure of their non-Markovian noise; both general and stemming from quantum correlations across time. We find that non-Markovian noise is present in all cases, with a significant proportion originating from genuine quantum correlations. A comparison with a simple system-environment model indicates that the influence of nearest-neighbour qubits is not sufficient to account for all the observed quantum non-Markovianity.

{Non-Markovian noise can arise from slowly drifting environmental fluctuations, such as from electronics, material defects and even other quantum systems, e.g., idling qubits on a superconducting chip. These are ubiquitous in current day systems and could become fidelity-limiting if not properly mitigated.} Our results demonstrate the capacity of our methods to fully describe multi-time processes {in any platform with mid-circuit measurement capabilities} and confidently measure and characterise non-Markovian noise.

{Finally, multi-time quantum process tomography bears similar limitations and challenges as ordinary quantum-state tomography, for example idealised characterisation of measurements and preparations, stability of the experimental platform across multiple experimental runs, and exponential complexity in the number of time steps/qubits. While such issues are only a minor concern in the demonstration we have presented, which involves small, well-characterised devices, they can become substantial roadblocks for larger-scale applications and will need to be addressed in future work. While some work has been done to adapt mitigating strategies from quantum state tomography to partial or restricted multi-time tomography \cite{White2025whatcanunitary, Li_2024}, it is an open direction to apply related techniques to complete tomography experiments, as presented here.}

\begin{acknowledgments}
We thank Xin He for providing support in preparing the tomography experiments on the $\texttt{uq}$ processor, and Gerardo Paz Silva for sharing relevant literature.
We acknowledge the use of IBM Quantum services for this work and IARPA and MIT Lincoln Laboratory for providing TWPA to use in the experiment. The views expressed are those of the authors and do not reflect the official policy or position of IBM or the IBM Quantum team.
CG acknowledges funding from the Sydney Quantum Academy Fellowship and the UTS Chancellor's Research Fellowship. FC was supported by the Wallenberg Initiative on Networks and Quantum Information (WINQ). This work was supported by the Australian Research Council (ARC) Centre of Excellence for Quantum Engineered Systems grant (CE 170100009). The authors have benefited from the activities of COST Action CA23115: Relativistic Quantum Information, funded  by COST (European Cooperation in Science and Technology). Finally, we acknowledge the traditional owners of the land on which the University of Queensland, the University of Technology Sydney, and Macquarie University are situated, the Turrbal and Jagera people, the Gadigal people of the Eora Nation, and Wallumattagal clan of the Dharug Nation.

\end{acknowledgments}
\appendix
\section*{Appendix: Device characteristics, errors and simulation}
\label{sec:Appendix}
\subsection*{Experimental devices}\label{sec:methodsdevice}
The primary device used was a superconducting quantum processor consisting of five flux-tunable transmons, fabricated by Quantware~\cite{quantware}. System characteristics of the transmon we performed multi-time tomography on (`System') and the dominant interacting transmon (`Memory') are provided in Table \ref{table:uqchars}. These characteristics include T1 and T2, the relaxation and dephasing lifetimes. The `System' qubit only has one nearest-neighbour coupling to the `Memory' qubit, which in turn is coupled to two other qubits. Charge lines and flux lines to each qubit permit XY control and frequency tunability respectively. A common feedline couples to individual dedicated resonators for each transmon qubit. Typical gate fidelities on this device are 99.5\%, while readout fidelity is around 93\%. 

The custom-built open-source library \texttt{sqdtoolz}~\cite{sqdtoolz} is used to orchestrate the array of high-precision instruments required to execute control protocols on our quantum processor. This includes the synchronised generation and timing of microwave-frequency pulsed signals for control and measurement, a continuous microwave-frequency pump signal for travelling wave parametric amplifier (TWPA)~\cite{Macklin2015} operation and a stable DC signal for frequency tuning purposes. 

\begin{table}[ht!]{
\centering
\begin{tabular}{|r||*{2}{c|}}
\hline
\small{Characteristic} &\small{System qubit}&\small{Memory qubit} \\\hline
\small{SPF} &\small{5.97 GHz}&\small{5.03 GHz}\\\hline
\small{$\omega$} &\small{5.11 GHz}&\small{5.03 GHz}\\\hline
\small{$g_{12}$} &\small{N/A}&\small{11 MHz}\\\hline
\small{$T_1$} &\small{14 $\mu$s}& \small{10 $\mu$s}\\\hline
\small{$T_2$} &\small{1.8 $\mu$s} & \small{6 $\mu$s} \\\hline
\end{tabular}\par
}
\bigskip
\caption{The characteristics of the major interacting transmon qubits in the \texttt{uq} processor. SPF is symmetry point frequency, $w$ is the operational frequency, $g_{12}$ is the coupling to the system, $T_1$ is relaxation time and $T_2$ is dephasing time.  The system qubit is operating at a transition frequency ($\omega$) far from its symmetry point frequency, which increases its sensitivity to flux noise and in turn reduces $T_2$.}
\label{table:uqchars}
\end{table}

Experiments are also performed on the cloud 7-qubit quantum processor \texttt{ibm\_perth}, as provided by \texttt{IBM Quantum}~\cite{IBMQ}. This system offers increased coherence times and gate fidelities at the expense of frequency control. Transmon characteristics can be found in Table \ref{table:ibmchars}. Gate fidelities are reported at 99.95\%, and readout fidelity at 97.5\%. \\

\begin{table}[ht!]{
\centering

\begin{tabular}{|r||*{2}{c|}}
\hline
\small{Characteristic} &\small{System qubit}&\small{Memory qubit} \\\hline
\small{$\omega$} &\small{5.16 GHz}&\small{4.98 GHz}\\\hline
\small{$g_{12}$} &\small{N/A}&\small{3 MHz}\\\hline
\small{$T_1$} &\small{207 $\mu$s}& \small{102 $\mu$s}\\\hline
\small{$T_2$} &\small{295 $\mu$s} & \small{113 $\mu$s} \\\hline
\end{tabular}\par
}
\bigskip
\caption{The characteristics of the major interacting transmon qubits in the IBM cloud processor. For this system, we have no control over frequency and therefore no control over interaction strength.}
\label{table:ibmchars}
\end{table}

\subsection*{Error estimation}

The credible intervals reported in Table 1 in the main text, were calculated {using a parametric bootstrapping method}. Bootstrapping a standard deviation for some metric $\hat{\theta}(\hat{X})$ relies on the assumption that the relative frequency of the measured $\hat{X}$ is representative of the underlying distribution which $\hat{X}$ is drawn from. In our case, $\hat{X}$ is a measured set of shot probabilities, which may look something like $\{$\texttt{00}: 0.7773, \texttt{10}: 0.1738, \texttt{01}: 0.0421, \texttt{11}: 0.0068$\}$, where, $\texttt{00}$ for example corresponds to the probability of getting the plus state in both measurements at $B$ and $C$. {We also address the measurement error.  A number $p$ in the Table~\ref{table:measurement_errors} means that with probability $p$ the measurement of the corresponding eigenstate is correctly reported as that eigenstate. So $1-p$ is the missclassification probability. For example a measurement of $(X_+,Y_-)$ would be reported as $(X_+,Y_+)$ with probability $0.964\times (1-0.912)=0.085$. For \texttt{ibm\_perth} the characteristics of the two measurements were the same.} We resample 8000 shots from {a combination of simulated measurement errors given in Table~\ref{table:measurement_errors} and a multinomial distribution using the observed relative frequencies}, over 1000 trials, creating a set of bootstrapped results $\hat{X}_i$. The simulated distribution of the metric of interest is then evaluated as $\hat{\theta}(\hat{X}_{i})$ and twice the standard deviation of this distribution is used to calculate the 95\% credible interval of the reported metric $\hat{\theta}(\hat{X})$. This technique is designed to account for both shot noise in the finite-sample estimates of observables {and to give an estimation of the impact of measurement errors.} It does not account for systematic error at the hardware level (e.g. gate miscalibration). Finally, we rely on bias correction in the bootstrapping process; the sample mean of the metric $\hat{\theta}(\hat{X}_i)$ was in general not equal to the measured $\hat{\theta}(\hat{X})$ (sometimes outside of the credible interval).

\begin{table}[ht]
\normalsize
\centering
\begin{tabular}{cccc} \toprule
    {} & {} & {$+$ state} & {$-$ state} \\ \midrule
    \texttt{uq} & $m_1$  & 0.964 & 0.914  \\
    & $m_2$  & 0.945 & 0.912  \\\midrule
    \texttt{ibm\_perth} & $m_1$\&$m_2$  & 0.976 & 0.973 \\ \bottomrule \\
\end{tabular}
\caption{{Estimated measurement error probabilities for the two measurements $m_1$ and $m_2$ at $B$ and $C$.}}
\label{table:measurement_errors}
\end{table}

\subsection*{Simulated process}
\begin{figure}[ht]
    \centering
    \includegraphics[scale=.12]{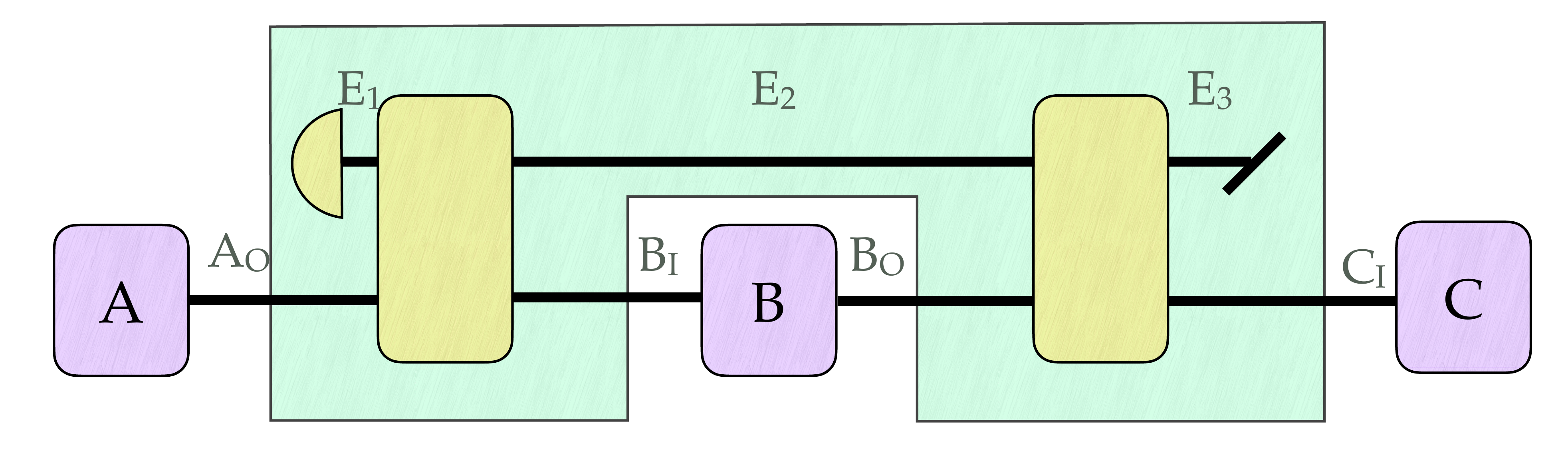}
    \caption{A three-time quantum process involving state preparation and two gates acting on both systems.}
    \label{fig:process_syst}
\end{figure}

In both platforms used, the qubit we probe belongs to a larger multi-qubit array. The particular qubits we choose each have one nearest-neighbour qubit, with which they are directly coupled, as well as additional distant qubits (two or more degrees of separation), with which interactions will be weaker. We therefore take the simplest model of our environment to be that of the single coupled qubit, as shown in Fig.~\ref{fig:process_syst}. This interaction can be described by the Hamiltonian~\cite{Blais2007}
\begin{equation}
    H = \frac{\omega_1}{2}(\id\otimes\sigma_z) + \frac{\omega_2}{2}(\sigma_z\otimes\id) + g_{12}(\sigma_+\otimes\sigma_- + \sigma_-\otimes\sigma_+),
    \label{eq:model}
\end{equation}
where $\omega_{1,2}$ are the resonance frequencies of the system and memory qubit, respectively, and $\sigma_{+,-}$ are the creation and annihilation operators, respectively. Then the free evolution of system-memory (the environment) is expressed as $U =e^{-iHt}$, where $t$ is the time of the evolution. 

To write the process matrix for our system, we start with writing the terms: the initial state of the environment  $\rho^{E_1}$, the channel $T_1^{A_OE_1^TB_IE2}$ and the second channel $T_2^{B_OE_2^TC_IE_3}$ (see Fig \ref{fig:process_syst}), where the superscript $T$ denotes partial transposition. To obtain the process matrix, we combine the terms with the link product and trace out all the environment systems 
\begin{equation}
    W = \text{Tr}_{E_1,E_2,E_3}(\rho^{E_1}* T_1^{A_OE_1^TB_IE_2}* T_2^{B_OE_2^TC_IE_3}),
\end{equation}
where $\rho^{E_1} = \ket{0}$ because all qubits on the chip are initialised to the ground state, $T_1^{A_OE_1B_IE2} = [[U]]^{A_OE_1B_IE2}$ and $T_2^{B_OE_2^TC_IE_3} = [[U]]^{B_OE_2C_IE_3}$, and $[[U]]$ is the Choi form of $U$, without the partial transposition.  

\emph{Choi matrix definition:} The Choi matrix $M^{A_IA_O}\in {\cal L}({\cal H}^{A_I}\otimes{\cal H}^{A_O})$, isomorphic to a CP map ${\cal M}^{A} : {\cal L}({\cal H}^{A_I}) \rightarrow {\cal L}({\cal H}^{A_O})$ is defined as $M^{A_IA_O} := [{\cal I} \otimes {\cal M}(\KetBra{\id}{\id})]^T$, where $\cal I$ is the identity map, $\Ket{\id} = \sum_{j=1}^{d_{A_I}}\ket{jj} \in {\cal H}^{A_I} \otimes {\cal H}^{A_I}$, $\{\ket{j}\}^{d_{A_I}}_{j=1}$ is an orthonormal basis on ${\cal H}^{A_I}$ and ${T}$ denotes matrix transposition in that basis and some basis of ${\cal H}^{A_O}$.

\emph{Link product definition:} $W^{xyz}= \rho^x*T^{yz} = \left[ \left(\rho^{x} \otimes \id^{yz} \right)  \left(\id^{x}\otimes T^{yz}\right)\right]$.

\begin{figure}[]
    \centering
    \begin{tikzpicture}[scale=1.4]
  \node[inner sep=10pt] at (3,4.6){\includegraphics[scale=0.5]{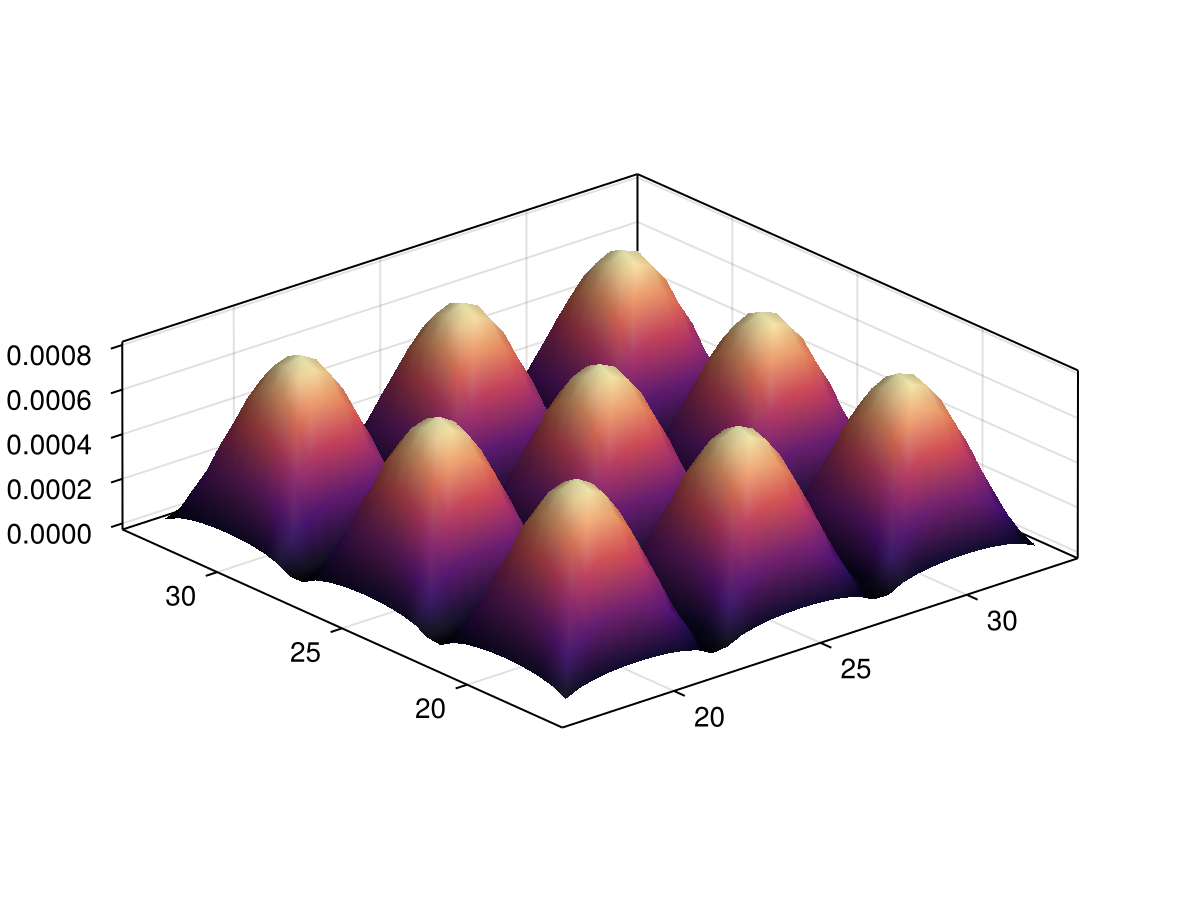}};
  \draw (0,4.6) node {\(\mathcal{D}\)};
  \node[inner sep=10pt] at (3,2){\includegraphics[scale=0.5]{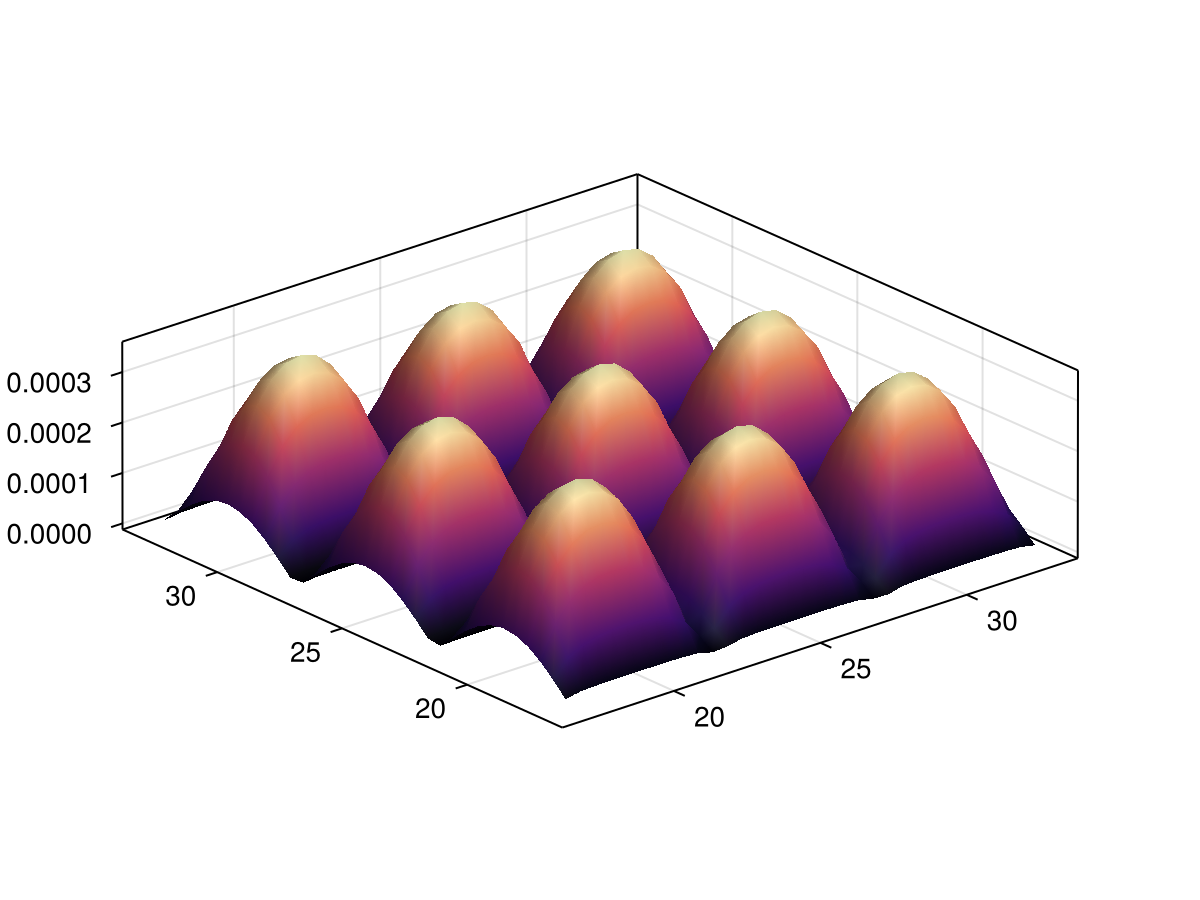}};
  \draw (0,2) node {\(\mathcal{N}\)};
  \draw (1.5,0.7) node {\(t_{1}\)};
  \draw (4.5,0.7) node {\(t_{2}\)};
  \draw (1,5.6) node {(a)};
  \draw (1,3.1) node {(b)};
    \end{tikzpicture}
    \caption{Simulated (a) Relative entropy and (b) Negativity values for the $\texttt{ibm\_perth}$ processor over a range of free evolution times $t_1$ and $t_2$.}
    \label{fig:ibm_sim}
\end{figure}

We see that the free evolution, U, depends on the parameters of the Hamiltonian, $\omega_1$, $\omega_2$, $g_{1,2}$, and the times $t_1$ and $t_2$ for which the interaction lasts. Hence, our process matrix will also depend on these parameters, $W(\omega_1,\omega_2, g_{1,2},t_1,t_2)$. To simulate the process on $\texttt{uq}$, experimental parameters were measured to be as given in Table~\ref{table:uqchars}, and we scan $t_1$ and $t_2$ from 85 to 115 ns, capturing the 97 ns process conducted in the experiment. In the experiment, we chose $t_1=t_2=$ 97 ns because in the simulation these times maximised non-Markovianity. For the process on $\texttt{ibm\_perth}$, the parameters were fixed at the values available in Table~\ref{table:ibmchars}, and we scan $t_1$ and $t_2$ between 17 and 33 ns, capturing the permutations of $t_1 = [21.33, 24.89, 28.44]$ ns and $t_2 = [21.33, 24.89, 28.44]$ ns conducted in the experiment{, which are centered around a peak of non-Markovianity in the simulation}. The $\texttt{ibm\_perth}$ process simulation yielded the results in Fig.~\ref{fig:ibm_sim}. We obtained a similar structure for the $\texttt{uq}$ process.

According to these results, the experimental runs where evolution times are $t_1=t_2= 24.89$ ns should maximise both metrics of non-Markovianity on $\texttt{ibm\_perth}$, while other settings ($t_1 = 21.33$ ns and $t_2 = 28.44$ ns) should be close to fully suppressing non-Markovian dynamics. In practice, both metrics were observed to be smeared across the free evolution times scanned. This is not a particularly surprising result when contrasting the simplicity of the coupled-qubit environment described in Eq.~\ref{eq:model} to the complexity of current-day quantum hardware, including but not limited to the extra qubits on each chip. 

\begin{table}[ht]
\normalsize
\centering
\begin{tabular}{cccc} \toprule
    {} & {} & {max$(\mathcal{D})$} & {max$({\mathcal{N}})$} \\ \midrule
    \texttt{uq} & measured  & 0.2301 & 0.0216  \\
    & simulated  & 0.0613 & 0.0229  \\\midrule
    \texttt{ibm\_perth} & measured  & 0.0560 & 0.0060 \\
    & simulated & 0.0007 & 0.0003 \\ \bottomrule \\
\end{tabular}
\caption{The maximum values observed for the relative entropy ($\mathcal{D}$) and negativity ($\mathcal{N}$) across the experimentally measured (across the runs of Table 1 of main text) and simulated processes.}
\label{table:sim_metrics}
\end{table}

A point of some interest is the comparison of the maximum $\mathcal{D}$ and $\mathcal{N}$ predicted in simulation (across a reasonable ($t_1$, $t_2$) range) to the values observed in the experiment. A summary of these is provided in Table~\ref{table:sim_metrics}. The qualitative structure of the simulated values matches the results presented in this work and are reasonable quantitatively on the $\texttt{uq}$ processor. Both metrics are heavily underestimated on the $\texttt{ibm\_perth}$ processor. This is not an unusual result; the Hamiltonian presented in Eq~\ref{eq:model} is certainly missing several terms, both characterisable (e.g. other qubits on the chip) and unknown. Hence, although the naive model is a reasonable first-order approximation, our results show that we can probe more complex dynamics than we can simulate, making our methods a powerful tool to detect non-Markovian noise. 

\bibliographystyle{quantum}
\providecommand{\noopsort}[1]{}\providecommand{\singleletter}[1]{#1}%

\end{document}